\documentstyle[epsfig]{mn} 
\input{epsf}   
   
\newcommand{\bez}{\begin{eqnarray*}} 
\newcommand{\eez}{\end{eqnarray*}} 
\newcommand{\be}{\begin{equation}} 
\newcommand{\ee}{\end{equation}} 
\newcommand{\beq}{\begin{eqnarray}}  
\newcommand{\eeq}{\end{eqnarray}}  
\newcommand{\bc}{\begin{center}}  
\newcommand{\ec}{\end{center}} 
   
\newbox\grsign \setbox\grsign=\hbox{$>$} \newdimen\grdimen \grdimen=\ht\grsign   
 
\newbox\simlessbox \newbox\simgreatbox \newbox\simpropbox   
\setbox\simgreatbox=\hbox{\raise.5ex\hbox{$>$}\llap   
     {\lower.5ex\hbox{$\sim$}}}\ht1=\grdimen\dp1=0pt   
\setbox\simlessbox=\hbox{\raise.5ex\hbox{$<$}\llap   
     {\lower.5ex\hbox{$\sim$}}}\ht2=\grdimen\dp2=0pt   
\setbox\simpropbox=\hbox{\raise.5ex\hbox{$\propto$}\llap   
     {\lower.5ex\hbox{$\sim$}}}\ht2=\grdimen\dp2=0pt   
\def\simgt{\mathrel{\copy\simgreatbox}}   
\def\simlt{\mathrel{\copy\simlessbox}}

\def\Tbb{T_{\rm bb}}   
\def\Te{T_{\rm e}}   
\def\tT{\tau_{\rm T}}   
\def\td{\tau_{\rm d}}   
\def\ta{\tau_{\rm a}}

\def\tB{\tau_{\rm B}}

\def\pesch{\dot{P_{\rm h}}}   
\def\pescr{\dot{P_{R}}}   
\def\pescs{\dot{P_{\rm s}}}   
\def\pescx{\dot{P_{\rm x}}}   
\def\pesc{\dot{P}}

\def\Ls{L_{\rm s}}   
\def\Lh{L_{\rm h}}   
\def\Luv{L_{\rm uv}}   
\def\Lx{L_{\rm x}}   
\def\LR{L_{R}}   
   
\def\Uh{U_{\rm h}}   
\def\Uuv{U_{\rm uv}}   
\def\Ux{U_{\rm x}}   
\def\UR{U_{R}}   
\def\aR{a_{R}}   
\begin{document}

\title[X--rays produced by a hot plasma containing cold clouds]{X-ray spectra produced by a hot plasma containing cold clouds}  
     
\author[J.~Malzac, A. Celotti]{\parbox[]{6.8in} {Julien~Malzac$^{1}$,  
Annalisa Celotti$^{2}$}\\  
$^{1}$Osservatorio Astronomico di Brera, via Brera, 28, 20121 Milan, Italy\\  
$^{2}$SISSA, via Beirut 2-4, 34014 Trieste, Italy}

\date{Accepted, Received}   
  
\maketitle


\begin{abstract} 
We compute the hard X-ray spectra  from a hot plasma pervaded by small 
cold  dense clouds.   The  main  cooling mechanism  of  the plasma  is 
Compton  cooling by  the soft  thermal emission  from the  clouds.  We 
compute numerically the equilibrium temperature of the plasma together 
with the escaping spectrum.  The spectrum depends mainly on the amount 
of cold clouds  filling the hot phase.  The  clouds covering factor is 
constrained to  be low  in order to  produce spectra similar  to those 
observed  in Seyfert  galaxies and  X-ray binaries,  implying  that an 
external reflector is required in order to reproduce the full range of 
observed reflection  amplitudes.  We also  derive analytical estimates 
for the X-ray spectral slope  and reflection amplitude using an escape 
probability formalism. 
\end{abstract}  
  
\begin{keywords}   
{accretion, accretion discs -- black hole physics -- radiative   
transfer -- gamma-rays: theory -- galaxies: Seyfert -- X-rays:   
general}   
\end{keywords}   
   

\footnotetext{E-mail: malzac@brera.mi.astro.it}  
  
\section{Introduction}  
  
The physical conditions in the inner parts of the accretion flow 
surrounding a black hole are likely to be very chaotic.  A situation 
that has been often considered in the literature is that of the 
so-called 'cauldron', where a soup formed by a hot plasma contains 
small grains constituted by small dense clouds of much colder matter 
(e.g. Guilbert \& Rees 1988).  Several works were devoted to explain 
how such a configuration could be physically realized and compute the 
spectrum emitted by the clouds for different cloud optical depths (see 
e.g.  Rees 1987; Ferland \& Rees 1988; Rees, Netzer \& Ferland 1989; 
Celotti, Fabian \& Rees 1992; Kuncic, Blackman \& Rees 1996; 
Collin-Souffrin et al. 1996; Kuncic, Celotti \& Rees 1997; Krolik 
1998).  In particular the main observable effect of optically thick 
cold clouds is the reprocessing into soft UV photons of the hard X-ray 
radiation produced in the hot phase by the Comptonisation process. 
This reprocessed emission is likely to contribute, at least partly, to 
the big blue bump observed in AGN.  In addition, the presence of such 
clumps inside the hot plasma may also contribute to the formation of a 
reflection component (see e.g. Nandra \& George 1994), responsible for 
the bump in the hard X-ray domain, commonly observed in Seyfert 
galaxies and galactic black hole candidates. 
 
Most of the previous works on  cloud models focused on the physics and 
radiative  processes  in the  clouds themselves,  without taking  into 
account the possible  effects of the clouds on  the characteristics of 
the  hot phase.   Indeed the  soft radiation  re--emitted by  the cold 
material can  constitute the main radiation field  responsible for the 
Compton cooling of the hot  gas.  This in turn affects the temperature 
of the hot plasma, and thus the emitted X-ray Comptonised spectrum. 
  
This feedback loop is identical to that found in accretion disc corona 
models (Haardt \& Maraschi 1993).  Similarly, the conditions at 
radiative equilibrium depend mainly on the cold matter distribution 
relative to the hot plasma. In this context, Malzac (2001) (hereafter 
M01) studied a geometry where the clouds are external to the hot 
Comptonising plasma and spherically distributed around it.  The aim of 
the present work is to study the effects of the presence of cold 
optically thick clouds distributed \emph{inside} the hot phase.  As 
demonstrated below, the cooling by the cold clouds is then more 
efficient than in the case of an external reprocessor. 
 
Under assumptions detailed in section~\ref{sec:assump}, we use a 
numerical approach based on non-linear Monte-Carlo simulations, 
described in section~\ref{sec:numap}, to compute the emitted spectra. 
These numerical results are found to be in agreement with analytical 
formulae, derived in section~\ref{sec:analrg}, giving the slope of the 
primary X-ray Comptonised spectrum $\Gamma$ and the reflection 
amplitude $R$.  The predictions of the model are then discussed and 
compared with the data in section~\ref{sec:results}. 
     
\section{Model assumptions and parameters} 
\label{sec:assump}  
 
\begin{figure} 
\scalebox{.5}{\includegraphics{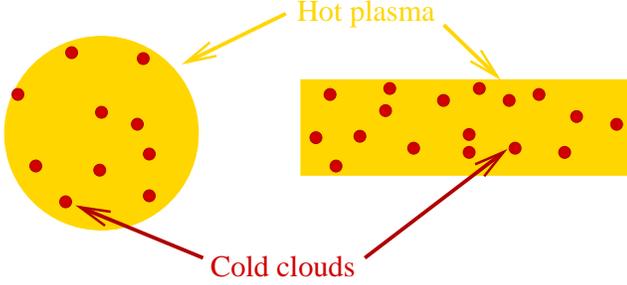}}\\   
\caption{The  hot plasma emitting  the hard  X- and  soft $\gamma$-ray 
emission  is  pervaded by  small  lumps  of  cold dense  matter.   The 
geometry of the hot plasma could  be close to spherical (left) or slab 
(right).  The  cold matter is assumed to  be homogeneously distributed 
inside the plasma volume.  The  ambient high energy radiation which is 
intercepted by  the cold  clouds is partly  reprocessed as  low energy 
(UV, EUV)  radiation and partly  reflected in the X-  and $\gamma$-ray 
energy   domains.   The  system   is  assumed   to  be   in  radiative 
equilibrium.}\label{fig:geo} 
\end{figure}

The physical characteristics of the clouds are of prime importance for 
the emitted spectrum. If they are optically (Thomson) thin, and with a 
large covering fraction, partial transmission of the X-ray radiation 
across a cloud is likely to produce strong absorption features in the 
X-ray spectrum (e.g.  Kuncic, Celotti \& Rees 1997).  On the other 
hand, if the clouds are optically thick and highly absorbing they do 
not produce, independent of their covering fraction, strong apparent 
absorption features, as indeed observed (Nandra \& George 1994). 
  
For this reason in this analysis we will consider the case of 
optically thick clouds with zero transmission, which corresponds to 
typical column densities $\simgt10^{25}$ cm$^{-2}$. We assume that the 
individual cold clouds are much smaller than the characteristic size 
of the region occupied by the hot plasma: if the emitting region has a 
typical dimension $H$, the typical cloud size is of order $\epsilon H$ 
with $\epsilon\simlt 10^{-2}$ \footnote{ For simplicity we will assume 
spherical clumps, although filamentary or sheet-like geometries might 
be also appropriate.}.  This constrains the cloud hydrogen density to 
be large, larger than $(\sigma_{\rm T}\epsilon H)^{-1}$.  This limit 
writes $n_{H}\simgt 10^{12}$ cm$^{-3}$ in the case of Seyfert galaxies 
(AGN, with $H\simeq 10^{15}$ $H_{15}$ cm), and $n_{H}\simgt 10^{20}$ 
cm$^{-3}$ in the case of galactic black holes (GBH, where $H \simeq 
10^{7}$ $H_7$ cm).  Such large densities in turn constrain the 
ionization parameter to be relatively low ($\xi$$<$10$^{3}$ for a 
$10^{45}$ erg s$^{-1}$ luminosity AGN, and $\xi$$<$10$^{2}$ for a 
$10^{36}$ erg s$^{-1}$ GBH) and thus the possible effects of 
ionization are likely to be weak (\.Zycki et al.  1994).  In our 
calculations we will then assume that the clouds are neutral. 
  
We will consider a system in radiative equilibrium.  The effective 
temperature of the clouds $k\Tbb$ can be simply estimated.  Assuming 
that the flux is uniform inside a spherical hot phase of bolometric 
luminosity $L$, a cloud surface element will receive a flux $F=L/(4 
\pi H^{2})$.  If the clouds are dense enough this flux is fully 
absorbed and re--radiated as a quasi-blackbody (e.g.  Ferland \& Rees 
1988) and the Stefan--Boltzmann's law gives well known characteristic 
energies: 
\begin{equation}  
k\Tbb \simeq  2.97 \frac{L_{45}^{1/4}}{H_{15}^{1/2}} \mathrm{eV}  
     \simeq 167 \frac{L_{36}^{1/4}} {H_{7}^{1/2}} \mathrm{eV},  
\label{eq:temp}  
\end{equation}  
where $L_{45}$ and $L_{36}$ are the bolometric luminosity expressed 
respectively in $10^{45}$ and $10^{36}$ erg s$^{-1}$ units. We will 
assume that the radiation reprocessed by the clouds is the only source 
of soft seed photons. 

As we are focusing on the spectral signatures of this two phase medium 
and as they do not constrain -- as already mentioned -- the physical 
status of the clouds (as long as they are optically thick and dense 
enough to maintain a low temperature), the clumps might exist with a 
large range of conditions.  Nonetheless the range of parameters 
considered (in dimension and density) are consistent with assuming 
that the gas reaches radiative equilibrium (i.e.  the cooling 
timescales are shorter than the dynamical one) and pressure 
equilibrium (with a size $\epsilon H$ small enough to reach it in a 
dynamical timescale). 
 
In general if neutral clouds are confined by a hot ionized environment 
the thermal pressure balance 
can be written in terms of the cloud Thomson optical 
depth $\tau=n_{H}\sigma_{T}\epsilon H$: 
\begin{equation}  
\tau \sim 2 \epsilon \tT T_{\rm e}/\Tbb.  
\label{eq:preseq}  
\end{equation}  

For typical observationally inferred values of $\tT=1$ and 
$kT_{e}$=100 keV, and using the expressions~(\ref{eq:temp}) for the 
temperature of the cold clouds, equation~(\ref{eq:preseq}) reads: 
\begin{equation}  
\tau \sim 64 \epsilon_{-2} \frac{H_{15}^{1/2}}{L_{45}^{1/4}}  
      \sim 12 \epsilon_{-2} \frac{H_{7}^{1/2}}{L_{36}^{1/4}},  
\end{equation}  
where $\epsilon= 10^{-2}\epsilon_{-2}$. We note however that this
simple estimate does not consider the possibility of a two temperature
plasma and the consequent effects of hot ions on the thermal balance
(see Zdziarski 1998; Spruit \& Haardt 2000; Deufel \& Spruit 2000). 

Thus the thick clouds considered here can be confined by the hot 
phase, but in any case thicker clouds may exist and be 
e.g. magnetically confined (for more details see e.g. Kuncic, Blackman 
\& Rees 1996).  Note that the clumps do not need to survive for 
longer that their cooling time provided they can be continuously 
replaced. 
 
Let us now  consider the spectrum produced in  these environment.  The 
three  main  parameters   controlling  the  resulting  X-/$\gamma$-ray 
spectral shape are the following: 
  
\begin{enumerate}  

\item The  amount of cold clouds  pervading the hot  plasma.  Since we 
consider small scale clumps we  can quantify the amount of cold matter 
by  using an  effective ``cloud''  optical  depth $\tB$,  such that  a 
photon  crossing the hot  plasma has  a probability  $1-\exp(-\tB)$ of 
intercepting a cloud. $\tB$ may  be thus defined using a cross section 
formalism, considering an individual cloud as a particle: 
\begin{equation} 
\tB=\frac{N}{V}\langle A \rangle H
\end{equation}
where $V$ is the plasma volume, $N$ is the total number of cold 
clouds, $\langle A \rangle$ is the average geometric cross section of each cloud. 
For spherical clouds pervading a spherical plasma, $\tB$ can be written explicitly in term of the number and size of the clouds: 
\begin{equation} 
\tB=\frac {3N\epsilon^{2}}{4}
\end{equation}

\item The Thomson optical depth $\tT$ of the hot plasma. 

\item The characteristic energy of the soft photons emitted by the 
clouds.  We assume a blackbody spectrum with a fixed temperature 
$k\Tbb$. As the radiative equilibrium is not extremely sensitive to 
the value of $k\Tbb$, we will fix $k\Tbb$=5 eV and $k\Tbb$=150 eV as 
typical values for Seyfert galaxies and GBH sources, respectively, as 
from equation~(\ref{eq:temp}). 
\end{enumerate}  
  
\begin{figure*}   
\scalebox{1}{\includegraphics{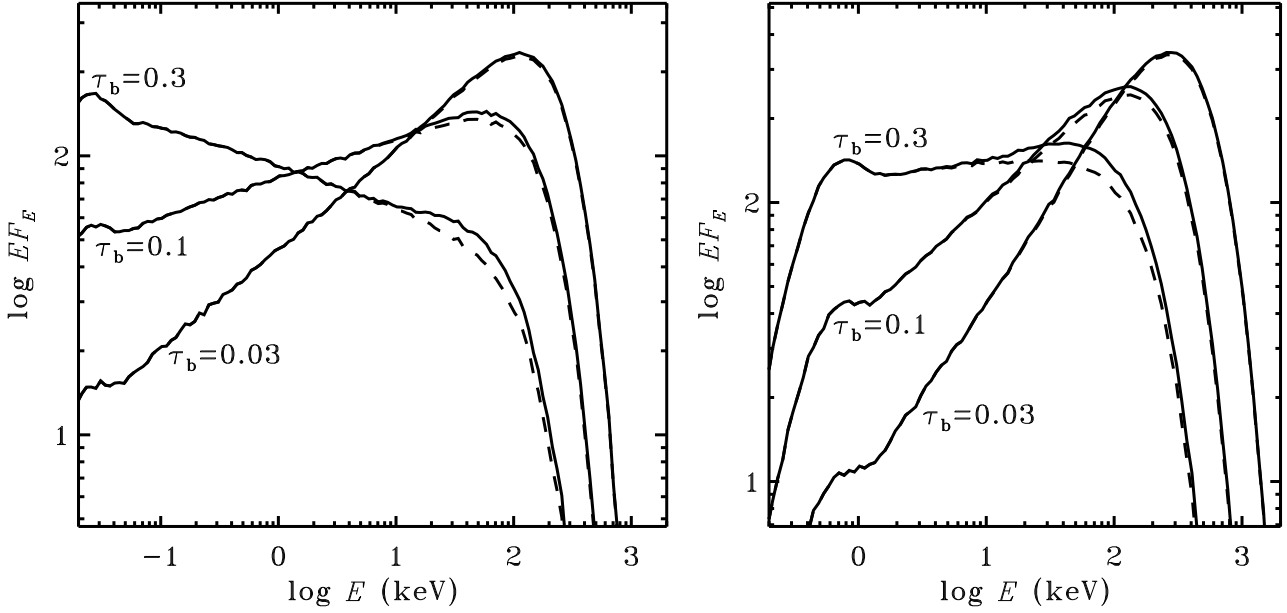}}   
\caption{Examples  of  spectra  produced  by  the lumpy  model  as  a 
function of the indicated values  of $\tB$.  The dashed lines show the 
resulting spectra  without the  reflection component.  The  left panel 
corresponds  to  the case  of  AGN  ($k\Tbb=5$  eV).  The  right  panel 
corresponds to  GBHs (k$\Tbb=150$ eV).  The spectra  (in arbitrary flux 
units)   are  angle-averaged   and  normalized   to  the   same  total 
luminosity. The plasma Thomson optical depth is $\tT=2$.} 
\label{fig:exspectra}   
\end{figure*}   
  
\begin{figure}  
\scalebox{1}{\includegraphics{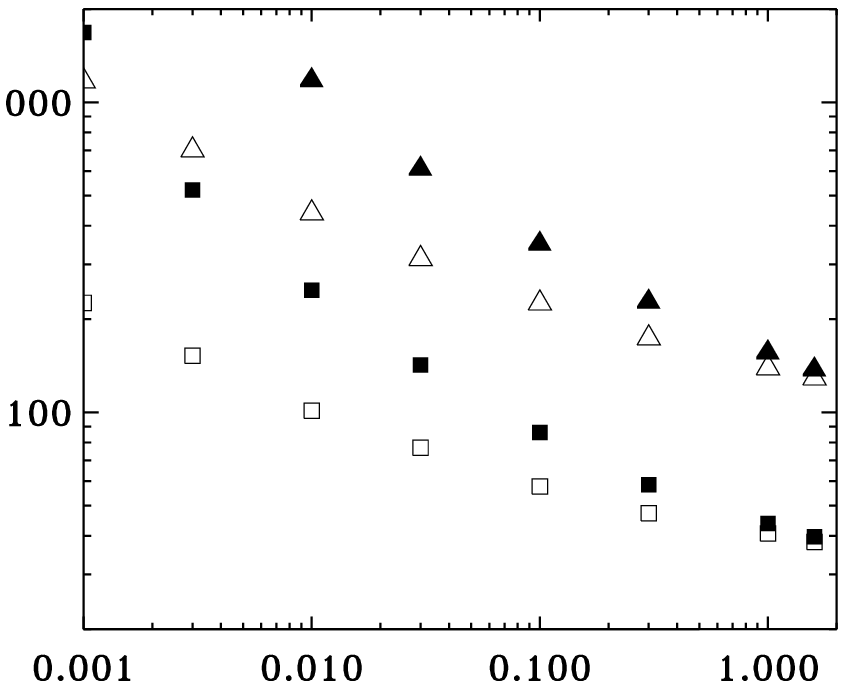}}  
\caption{   The   volume    averaged   electron   temperature   versus 
$\tB$. Filled and open symbols correspond to $k\Tbb=150$~eV (GBHs) and 
$k\Tbb=5$~eV (AGN),  respectively. Squares and triangles  refer to the 
cases $\tT=2$ and $\tT=0.5$.} 
\label{fig:tedetaub}  
\end{figure}  
 
As said before, the amount of cold matter and its distribution with 
respect to the hot plasma are of prime importance.  On the other hand, 
the actual geometry of the emitting region has little effects on the 
emitted spectrum.  We studied both the spherical and infinite slab 
geometries illustrated in Fig.~\ref{fig:geo}. As the results were 
qualitatively similar for both geometries, in the following we 
consider the spherical case only.

\section{Numerical approach} \label{sec:numap} 
  
In order to estimate the resulting physical parameters and spectra we 
use the Monte-Carlo code of Malzac \& Jourdain (2000) which is based 
on the Non-Linear Monte-Carlo method (NLMC) proposed by Stern (Stern 1985; 
Stern et al. 1995). We consider a spherical geometry.  The sphere is 
divided in 5 shells with equal thicknesses and is assumed to have a 
uniform density.  Both
$\tB$ and $\tT$ are defined along the radius $H$ of the sphere.

In the hot phase, the simulations are fully non-linear: both photons
 and electrons are followed using the Large Particles (LP) formalism 
described in Stern et al. (1995). 
This enables the equilibrium temperature to be computed 
in each zone according to the local energy balance heating = Compton cooling. 

The interaction  between a photon  and a cloud  is dealt as  a regular
Monte-Carlo interaction between two particles. However, 
as the cold clouds structure is assumed to be unaffected
 by the radiation field, the clouds themselves 
are not followed.
As the clouds are small and homogeneously distributed,
 $\tB$  is simply the  reaction rate for  1 photon,
with time expressed  in units of the light crossing  time $H/c$ of the
medium.  When  a LP photon  interacts with a  cloud, it is  assumed to
enter a semi-infinite slab medium with standard abundances and neutral
matter (Morrison  \& Mc~Cammon 1983).  The linear  Monte-Carlo code of
Malzac et  al.~(1998), is then used  to track the path  and the energy
changes of the  LP photon in this medium, until  it is either absorbed
or  escapes from  the cold  matter slab.   As long  as the  clouds are
optically thick,  the reflection spectrum  on a cloud does  not differ
much from  slab reflection (Nandra  \& George 1994).  A  more detailed
modeling of the  reflection component would rely on  a specific cloud
geometry which is essentially unknown.

The energy deposited in the cloud is re-injected in the hot phase in 
the form of thermal soft photons LP.  Both the reflected and 
reprocessed LP have a direction of propagation drawn from an isotropic 
distribution. 
 
\begin{figure*}  
\scalebox{1}{\includegraphics{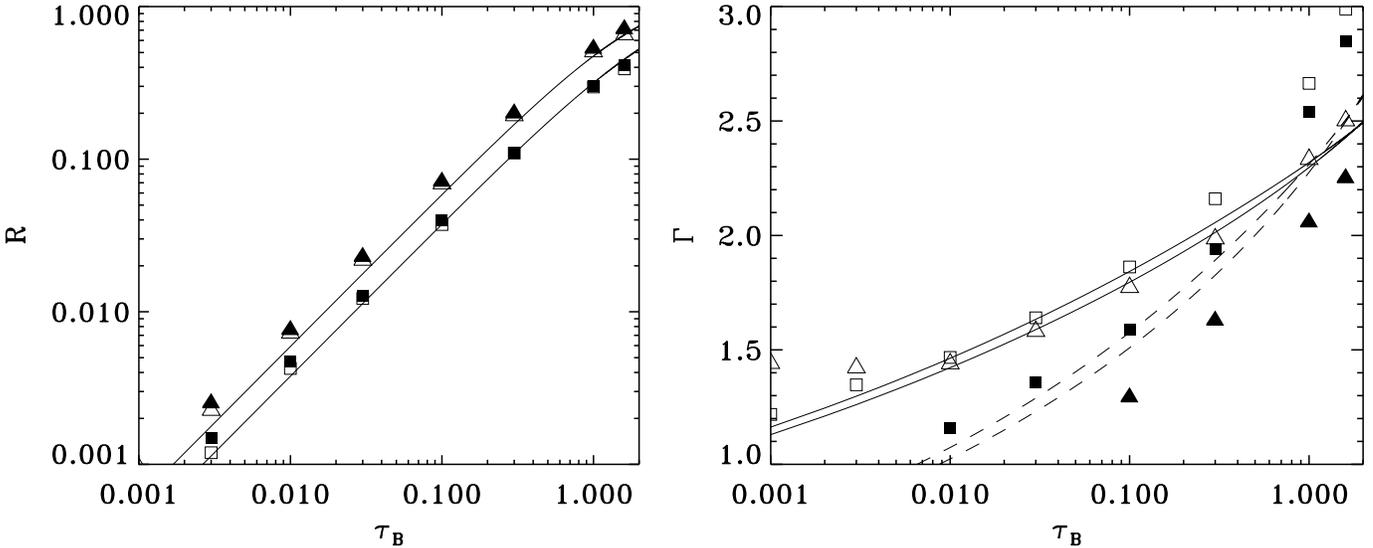}}\\  
\caption{Angle averaged reflection (left-hand panel) and spectral
index (right-hand panel) versus $\tB$. Filled and open symbols
correspond to $k\Tbb=150$~eV (GBHs) and $k\Tbb=5$~eV (AGN),
respectively.  Squares and triangles show the cases $\tT=2$ and
$\tT=0.5$.   In both panels, the curves represent the 
analytical approximation to the numerical results (see text). The
solid lines stand for $k\Tbb=5$~eV (AGN), the dashed lines for
$k\Tbb=150$~eV.}\label{fig:rgdetaub}
\end{figure*}

\section{Analytical estimates of $R$ and $\Gamma$}\label{sec:analrg}  
 
There are two parameters generally used to phenomenologically describe
the X-ray spectral properties of accreting black hole sources, namely
the amplitude $R$ of the reflection component with respect to the
primary continuum and the photon spectral index $\Gamma$ of the
power-law shaped primary component (see Malzac, Beloborodov \&
Poutanen 2001, MBP hereafter).

These parameters were determined from the numerical simulations in a
way similar to that described in MBP. The reflection was estimated by
computing the flux ratio between our simulated reflection, and the
angle-averaged slab-reflection produced by a source having the same
primary spectrum as the simulated one.
 
However, using the simple escape probability formalism described in 
the following, it is also possible to independently derive analytical 
estimates for $R$ and $\Gamma$, as follows.

A power $\Lh$ is deposited in the plasma as heating and escapes the  
system as three different kinds of radiation:  
\begin{equation} 
\Lh=\Luv+\Lx+\LR, 
\label{eq:lumbal} 
\end{equation} 
respectively the UV/soft X-ray reprocessed luminosity, the X-ray 
(Comptonized) luminosity, and the reflected luminosity.  Let $\Uh$ be 
the total radiative energy inside the hot plasma.  With similar 
notations as in equation~(\ref{eq:lumbal}) we have: 
\begin{equation} 
\Uh=\Uuv+\Ux+\UR.  
\label{eq:inten}  
\end{equation}  
  
The escaping luminosity is related to the internal energy through the 
escape probability (Lightman \& Zdziarski 1987): 
\begin{equation}   
\pesch=(\Lh/\Uh)(H/c),  
\label{eq:pesc}   
\end{equation}   
which depends on the geometry, sources distribution, optical depth and 
energy.  We consistently indicate as $\pescx$, $\pescs$,$\pescr$ the 
escape probability for Comptonised, soft and reflected radiation and 
will use the analytical approximation to $\dot P$ given in the 
Appendix (equation~(\ref{eq:peschomsph})). 
   
Let us now estimate the three different luminosities.    
  
The soft luminosity in the hot phase is produced by the clouds through 
absorption and reprocessing of the Comptonized and reflected radiation 
and disappears through escape or Comptonisation.  Thus, the radiative 
equilibrium balance for $\Uuv$ reads: 
\begin{equation}   
(\pescs+\tT)\Uuv= \tB(1-a)\Ux+\tB(1-\aR )\UR,   
\label{eq:usbal}   
\end{equation}   
where $a$ and $\aR$ are  the clouds energy and angle integrated albedo 
for a Comptonised and a reflection spectrum, respectively. For neutral 
matter  the typical  values are  $a\sim0.1$ (Magdziarz  \& Zdziarski 
1995; MBP) and $\aR\sim0.4$ (Malzac 2001). 
   
Similarly, the reflected radiation is formed through reflection of the 
Comptonised and reflected radiation on the clouds, and disappears via 
escape, Comptonisation in the hot plasma and absorption by the clouds: 
\begin{equation}   
(\pescr+\tT+\tB(1-\aR))\UR= \tB a \Ux.   
\label{eq:urbal}   
\end{equation}  
  
The fraction of the internal energy in the form of Comptonised 
radiation has as main source the power dissipated in the hot plasma 
but also the Comptonised soft and reflected radiation; the sink term 
is due to escape, absorption and reflection by the clouds: 
\begin{equation}  
(\pescx+\tB)\Ux= \Lh H/c+\tT \Uuv+\tT \UR.   
\label{eq:uxbal}  
\end{equation}  
  
Excluding the reflection component, the total luminosity sinking out 
of the hot phase either through escape or interaction with the clouds 
is: 
\begin{equation}  
L=(\pescs+\tB)\Ux +(\pescs+\tB)\Uuv,\label{eq:L}   
\end{equation}  
while the soft luminosity entering the hot phase (coming from the clouds)   
is given by equation~(\ref{eq:usbal}):   
\begin{equation}  
\Ls=(\pescx+\tT+\tB)\Uuv. \label{eq:Ls}    
\end{equation}     
  
Therefore combining equations (\ref{eq:L}) and (\ref{eq:Ls}) with 
equations (\ref{eq:usbal}) and (\ref{eq:urbal}) provides an estimate for 
the amplification factor $A\equiv L/\Ls$: 
\begin{equation}   
A=\frac{(\pescs+\tB)\tB+(\pescx+\tB)(\pescs+\tT)g}{\tB(\pescs+\tT+\tB)},   
\label{eq:ampli}   
\end{equation}   
where the factor $g$ is given by:   
\begin{equation}   
g=\frac{\pescr+\tT+\tB(1-\aR)}{(\pescr+\tT)(1-a)+\tB(1-\aR)}.   
\end{equation}   
Note that at first order $g\simeq1$ and $\pescx\simeq\pescs$ and thus:   
\begin{equation}   
A\sim 1+\pescx/\tB.   
\label{eq:ampli2}   
\end{equation}   
  
Finally the amplification factor is clearly directly related to the 
photon index $\Gamma$, which can therefore be expressed -- using the 
phenomenological formula given by Beloborodov (1999) -- as 
\begin{equation}   
\Gamma=2.33(A-1)^{-1/\delta}, 
\label{eq:belo}   
\end{equation}   
where the parameter $\delta=1/10$ for AGN and $1/6$ for GBHs. 
 
In addition equation~(\ref{eq:urbal}) provides an estimate for the 
second parameter, the reflection coefficient: 
\begin{equation} 
R\sim \frac{\LR}{a\Lx}  
\sim \frac{\pescr}{\pescx} \frac{\tB}{\pescr+\tT+\tB(1-\aR)}. 
\label{eq:R} 
\end{equation} 
This formula corresponds to an angle averaged reference 
slab-reflection spectrum.  Actually, the reflection coefficient also 
depends on the assumed inclination angle for the infinite slab model 
(see e.g. {\sc pexrav} model in {\sc XSPEC}, Magdziarz \& Zdziarski 
1995).  Equation~(\ref{eq:R}) could be corrected for this, for example 
by dividing it by the angular factor given by equation 2 of 
Ghisellini, Haardt \& Matt (1994).  However, for the relatively low 
inclination angles usually assumed in spectral fits, the correction is 
small and, for simplicity, we will neglect it.

\section{Results} \label{sec:results}  
 
Figure~\ref{fig:exspectra}  shows some  of the  simulated  spectra for 
different  values of  $\tB$.  The  radiative cooling  from  the clouds 
appears  to be  extremely  efficient and  starts  being effective  for 
values of $\tB$  as low as a few $\times 10^{-2}$.   At fixed $\tT$ and 
$\Tbb$, increasing $\tB$ increases  the cooling of the plasma (because 
of the enhancement of the  reprocessing and the subsequent increase of 
the soft radiation field) and therefore decreases its temperature (see 
below).  This affects the shape  of the emitted spectrum which becomes 
softer and cuts off at lower  energies.  Note that for the same set of 
parameters the spectra of GBHs  (right panel of the figure) are harder 
than the AGNs ones (left panel)  due to the larger energy domain where 
the   luminosity  can  be   released  in   the  latter   objects  (see 
e.g. MBP). The spectra shown are rather similar to what is observed in 
Seyfert  galaxies and  Galactic black  hole candidates  in  their hard 
state (see e.g. Zdziarski et al. 1997, Poutanen 1998). 
 
The dependence of  the hot plasma temperature on  $\tB$ is illustrated 
more   precisely  by  Fig.~\ref{fig:tedetaub}.    As  just   said  the 
temperature is generally lower in AGNs  (for the same $\tB$) and it is 
higher for  lower Thomson optically  depths.  While these  are general 
features  (see  e.g.   MBP)  of the  self-regulated  Compton  emission 
generally invoked  by the two-phase models (Haardt  \& Maraschi 1993), 
note that  in our case  the product $\Te\tT$  is $not$ expected  to be 
constant at fixed $\tB$ since  here the Compton parameter depends also 
slightly on $\tT$ (see Section~\ref{sec:analrg}). 
   
The observed range of temperatures -- say 50-200 keV -- inferred from 
observations of the high energy cut off in Seyfert galaxies and GBHs 
can be reproduced for $\tB$ in the range $10^{-2}$--$1$, depending 
also widely on $\tT$. In the optically thick case larger values of 
$\tB$ may induce very low temperatures. 
   
A  second observable  parameter is  the index  of the  (primary) X-ray 
spectrum. Fig.~\ref{fig:rgdetaub} shows the  evolution of the 2--10 keV 
spectral slope for increasing  $\tB$: as expected the spectrum softens 
very quickly for larger $\tB$. One  can see that the range of spectral 
slopes  observed  in  Seyfert  galaxies  and GBH  in  the  hard  state 
(approximately the range 1.4--2.2) can be achieved for $\tB$ roughly in 
the interval 10$^{-3}$--$1$  (larger $\tB$ clearly lead to  a too soft 
spectrum).  Note that  although  such interval  is  rather large,  the 
dependence on $\tB$ is quite steep especially in the GBH case. 
 
Finally let us examine the results on the intensity of the reflection 
component, shown in Fig.~\ref{fig:rgdetaub} (left-hand panel).  This 
immediately reveals that the inferred values of $R$ are rather low, 
even when $\tB$ approaches unity, and in general significantly lower 
than observed. According to equation~(\ref{eq:R}), $R$ values of order 
of unity can be achieved in the limit of large $\tB$. However, this 
corresponds to very large (and unobserved) $\Gamma$, as discussed in 
the next section. 
 
As shown in the same Fig.~\ref{fig:rgdetaub}, we also stress here that 
there is good agreement between the numerical results and the 
analytical estimates of $R$ and $\Gamma$ derived in section 
\ref{sec:analrg}.  Concerning $\Gamma$,we note however a significant 
discrepancy for the GBH optically thin case, for which the simulations 
give a much harder spectrum than the analytical estimates. Indeed, in 
this case the first Compton scattering order falls in the 2--10 keV 
range. This forms a bump making the spectrum locally harder in the 
energy range used to estimate $\Gamma$, and with a break at higher 
energy. This (unobserved) radiative transfer effect is due to the low 
optical depth ($\tT$=0.5) considered. A part from this the main 
discrepancies appear in the optically thick case and at large $\tB$ 
(corresponding to $\Gamma$ outside the observed range) arising because 
in this interval of parameters associated with low temperatures,
equation~(\ref{eq:belo}), which is formally valid for a spherical plasma 
with temperature in the range 50--100 keV, does not provide a good 
approximation to the $\Gamma$ vs $A$ relation anymore. 
  
\section{CONCLUSIONS}  
   
\begin{figure} 
\scalebox{1}{\includegraphics{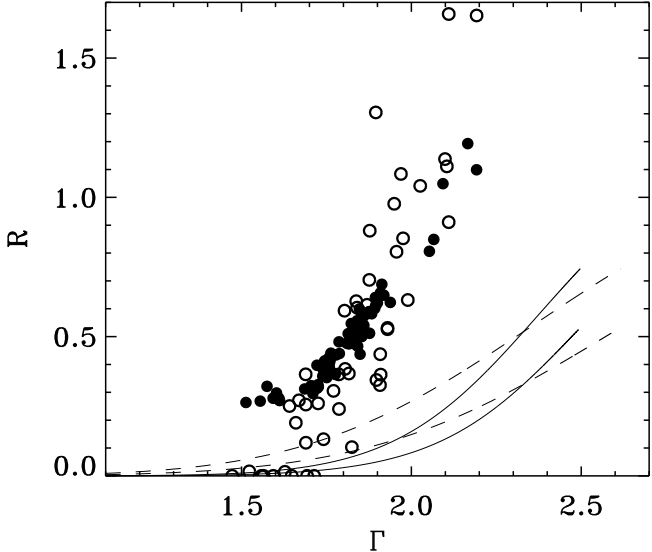}}\\   
\caption{ The circles show data in the $R$-$\Gamma$ plane 
corresponding to the sample of Seyfert galaxies (open circles) of 
Zdziarski et al.  (1999) and the sample of galactic black holes 
(filled circles) presented in Gilfanov et al. (2000).  The lines 
refer to the analytical model predictions: along these lines $\tB$ is 
varied at constant $\tT$.  Dashed and solid lines correspond to 
$k\Tbb$=$150$~eV (GBHs) and $k\Tbb$=$5$~eV (AGN) respectively. For 
both types of sources, the model predictions are plotted for $\tT$=$0.5$ 
(upper curves), and $\tT$=$2$ (lower curves), $\tB$ covers the range 0--2 along the curves.} 
\label{fig:rgcor} 
\end{figure} 
 
From the computation of the spectra expected in a scenario where cold 
reflecting and reprocessing gas is embedded in a coronal hot plasma we 
have inferred constraints on the distribution of such cold gas. In 
fact, having assumed that the gas can be described as small optically 
thick clouds, the main parameter regulating the amount of reprocessing 
(in turn responsible for the Compton cooling) and reflection is the 
cloud optical depth $\tB$, which quantifies the covering factor of the 
cold component.
 
This has to be relatively low. In fact, the observed range of spectral 
slopes constrains $\tB<1$.  This limit however forbids to produce 
simultaneously a reflection amplitude larger than $\sim0.3$.  The
situation is even more critical since $R\sim0.3$ is predicted only for
the softest sources which observationally appear to show the largest
reflection amplitude ($R\simgt1$), as shown in Fig.~\ref{fig:rgcor}.
 
Thus, in order to reproduce the spectral characteristics of numerous
objects (both Seyfert galaxies and GBH sources) with significant
reflection the model requires an additional reflecting medium which
does not contribute to the soft photon field, such as e.g. a cold
outer accretion disc, reflection on a torus in the case of Seyfert
galaxies or on the companion star in the case of X-ray binaries.

The reported correlation between $R$ and $\Gamma$ (shown in
Fig.~\ref{fig:rgcor}) is generally interpreted as being due to plasma
cooling on soft photons emitted by the same medium that gives rise to
the reflection component (see Zdziarski et al.  1999; Gilfanov et
al. 2000).  If the $R$-$\Gamma$ correlation were indicative of a
common medium for soft photon creation and reflection, such a medium
cannot be provided by cold matter mixed together with the hot phase,
and another reflector should contribute to, and even dominate the
radiative cooling of the hot plasma. 

A related problem is the observed correlation between the
degree of relativistic broadening of the Fe line with $\Gamma$ (Gilfanov et al. 2000
for black hole binaries, Lubi\'nski \& Zdziarski 2001 for Seyferts, see
however Yaqoob et al. 2002).  In the context of the present model this
correlation would suggest that the overall cloud distribution is not
homogeneous. The changes in the amount of cold material pervading the
hot plasma should be stronger in the innermost parts of the accretion
flow where the relativistic broadening is important.  However, since
the reflection amplitude produced by the cold clump is weaker than
observed, this requires that the bulk of the line would be emitted by
the cold clouds while the reflection would come mainly from an
external reflector.
 
Thus, the origin of
both soft photons and reflection are unlikely to be matter mixed
inside the hot Comptonising plasma.
In the framework of the cloud models this suggests that the cold
clouds should be essentially external to the hot Comptonising
plasma. Then the radiative cooling is much less efficient and the
reflection component may be larger.  Variations in the covering factor
and the distribution of the clouds may even produce a correlation
between $R$ and $\Gamma$ (as e.g. in the spherical case considered by
M01).

These conclusions should be however tempered by the fact that, 
in Seyfert galaxies, the $R$-$\Gamma$ correlation is still a
controversial matter on several grounds. 
Indeed, the measurment of the reflection amplitude $R$ 
suffers from large uncertainties.   
In particular, it appears that the $R$  values
 derived from spectral fits are very sensitive to
 the modelling of the primary spectrum 
(see e.g. Petrucci et al. 2001; Perola et al. 2002; Malzac \& Petrucci 2002) 
 as well as to the eventual presence 
of a secondary components such as a soft excess (Matt 2001). 
Possible effects such as ionization of the reflecting material,
 eventual relativistic smearing, Comptonisation of the reflected
 component may all affect the $R$ values derived
 from the neutral slab reflection model generally used ({\sc pexrav}).
We also note that in NGC 5506, Lamer, Uttley \& McHardy (2000) find an
anti-correlation between $R$ and $\Gamma$.

Also, even if the
correlation is real, its interpretation could be different.  Indeed,
following Nandra et al. (2000), Malzac \& Petrucci (2001, 2002) show that
at least part of the observed correlation
 could be understood in term of the effects of time-delays between
the fluctuations of the primary emission and the response of a
large-scale distant reflector (e.g a torus). Obviously, because of the time-scales involved, this interpretation cannot be valid 
for  black-hole binaries. On the other hand,
 in Seyfert galaxies, it is
supported by spectral and variability studies indicating that, in
several sources, an important fraction of the reflected features is
produced at large distances from the central engine (e.g.  Chiang et
al. 2000 for NGC 5548; Matt et al. 2001 and Lamer et al. 2000, for NGC
5506; Done, Madejski \& \.Zycki 2000 for IC4329a, Papadakis et al. 2002, for a sample of sources).  In a context
where the reflection features would be produced mainly on distant
material, the ``cauldron'' scenario considered here would remain a
viable possibility.
 
Future observational studies of the $R$-$\Gamma$  relation in samples
of sources as well as in individual objects should clarify this point.

\section*{Acknowledgments}
JM acknowledges grants from the Italian MURST
(COFIN98-02-15-41) and the European Commission (contract
number ERBFMRX-CT98-0195, TMR network "Accretion onto black holes,
compact stars and protostars"). AC acknowledges the MUIR for fundings.
We thank Andrzej~Zdziarski and Marat~Gilfanov
who provided the data used in this work.

\section*{APPENDIX: Fitting formulae for the escape probability}   
  
Let  us consider  a  medium  where both scattering and absorption  of 
radiation are  effective, with  $\td$ the total  optical depth  of the 
medium for the scattering process  and $\ta$ for absorption. Assume one 
injects instantaneously in the volume a number of photons according to 
a  given spatial  distribution.  The  photons can  disappear  from the 
medium either by real escape  (i.e. by crossing the volume boundaries) 
or by absorption.  Let $f$ be the distribution of the photons at their 
disappearance time, and $E$ the  fraction of photons that truly escape 
the  medium (i.e.  not absorbed).   Then, from  the definition  of the 
escape   probability   given    by   equation~(\ref{eq:pesc}),   it   is 
straightforward to show that: 
\begin{equation}  
\frac{1}{\pesc}=\frac{1}{E}\int_{0}^{+\infty}\quad dt \quad \left(1-\int_{0}^{t}f(t')dt'\right)\quad 
\end{equation}  
  
Using  a standard  Monte-Carlo method,  we estimated  $E$ and  the $f$ 
function for  different geometries and source  distributions. In these 
calculations  the  scattering process  was  assumed  isotropic and  the 
distribution  of scatterers  and absorbers  was homogeneous.   We then 
computed  numerically  $\pesc$ and  searched  for analytical  formulae 
giving  a reasonable representation  of the  numerical results  in the 
range $\td  <10$ and  $\ta < 10$.   The expressions given  below agree 
within  5$\%$  with  our  numerical  results.   For  slab  and  sphere 
geometries, the optical  depths $\ta$ and $\td$ are  defined along the 
full height of the slab and the radius of the sphere, respectively. 
 
\noindent For a spherical geometry and uniform isotropic injection:  
\begin{equation}  
\frac{1}{\pesc}=0.75\left[1+\frac{0.25\td}{1+0.49\ta(1+0.1\td)}\right](1+2.28\ta)^{0.16}.\label{eq:peschomsph}  
\end{equation}  
\noindent In the  limit $\ta$=0 this formula reduces  to that proposed 
by Stern et al. (1995). 
   
\noindent For a spherical geometry and central isotropic injection:  
  
\begin{eqnarray}  
\frac{1}{\pesc}&=&\{1.+0.45\td[1.+ 3.384\,10^{-3}\ta^{1.168}\nonumber\\  
               &&\exp\{3.3(1.09-0.10887\ta)(10.-\td)^{8.5\,10^{-3}}\nonumber\\  
               && +0.2999\ta+(0.346+0.1\td)\td^{0.681}\}]\}\nonumber\\  
               && \frac{\exp{(0.944\ta+1.32\,10^{-3}\ta^{2})}}{1+0.56\ta}.  
\end{eqnarray}

\noindent For an infinite  slab geometry and homogeneous and isotropic 
source function: 
\begin{eqnarray}  
\frac{1}{\pesc}=1&+&0.392\left(\frac{183.7\ta}{1+183.6\ta}\right)^{533}\nonumber\\  
&+&\frac{\td/3}{(1+0.08\td^{0.57}) (1+0.393\ta^{1.045})}.  
\end{eqnarray}

\noindent For  an infinite slab geometry and
photons injected isotropically at midplane:
\begin{eqnarray} 
\frac{1}{\pesc}&=&0.992\frac{1+123.2\ta}{1+124.9\ta}\exp[0.250\ta^{1.211}\nonumber\\ 
                &&+0.401\td^{0.602+(0.748\ta-0.0464\ta^{2})^{0.813}}].  
\end{eqnarray}

\noindent  These expressions  are general  and can  be applied  to any
situation where both absorption  and scattering are important.  For the
situation considered in this  work, the scattering $\td$ and absorption
$\ta$ optical  depths depend on  the type of radiation  considered (see
section~\ref{sec:analrg}):
\begin{itemize}
\item For the Comptonised component:
$\td$=$\tT$  and   $\ta$=$\tB$
\item For the reflected component: $\td$=$\aR\tB$, $\ta$=$\tT$+$($1-$\aR$)$\tB$ 
\item For the  soft component: $\td$=$\tB$,  $\ta$=$\tT$ 
\end{itemize}
Note  that we
neglect  the  relativistic  effects   due  to  the  reduction  of  the
Klein-Nishina cross section at high energies.
\end{document}